\documentclass[twocolumn]{jpsj2}
\usepackage{graphicx}
\usepackage{amsmath,amssymb}

\def\runauthor{Youichi {\sc Yanase} and Naoyuki Yorozu}

\title{ 
Localization and Superconductivity in Doped Semiconductors
} 

\author{Youichi {\sc Yanase}\footnote{E-mail:
yanase@hosi.phys.s.u-tokyo.ac.jp} and Naoyuki Yorozu}

\inst{Department of Physics, University of Tokyo, Tokyo 113-0033, Japan}

\recdate{14. Oct. 2008}

\abst
{ 
Motivated by the discovery of superconductivity in boron-doped (B-doped) diamond, 
we investigate the localization and superconductivity in heavily doped 
semiconductors. 
The competition between Anderson localization and 
$s$-wave superconductivity is investigated 
from the microscopic point of view. 
 The effect of microscopic inhomogeneity and the thermal fluctuation 
in superconductivity are taken into account using the self-consistent 
1-loop-order theory with respect to superconducting fluctuation. 
 The crossover from superconductivity in the host band to that 
in the impurity band is described on the basis of the disordered three-dimensional 
attractive Hubbard model for binary alloys. 
 We show that superconductor-insulator transition (SIT) accompanies 
the crossover. 
 We point out an enhancement of Cooper pairing in the crossover regime. 
 Further localization of the electron wave function gives rise to 
incoherent Cooper pairs and the pseudogap above $T_{\rm c}$. 
 A global phase diagram is drawn for host band superconductivity, 
impurity band superconductivity, Anderson localization, Fermi liquid state, 
and pseudogap state. 
 A theoretical interpretation is proposed for superconductivity in 
the doped diamond, SiC, and Si. 
}

\kword
{
Anderson localization; superconductor-insulator-transition; 
localized Cooper pairs; heavily doped semiconductors
}

\begin{document}
\sloppy
\maketitle

\newcommand{\con}{cond-mat/}
\newcommand{\eli}{$\acute{{\rm E}}$liashberg }
\renewcommand{\k}{\vec{k}}
\newcommand{\kk}{\vec{k'}}
\newcommand{\q}{\vec{q}}
\newcommand{\Q}{\vec{Q}}
\renewcommand{\r}{\vec{r}}
\newcommand{\e}{\varepsilon}
\newcommand{\ee}{\varepsilon^{'}}
\newcommand{\s}{{\mit{\it \Sigma}}}
\newcommand{\J}{\mbox{\boldmath$J$}}
\newcommand{\vv}{\mbox{\boldmath$v$}}
\newcommand{\Jh}{J_{{\rm H}}}
\newcommand{\LL}{\mbox{\boldmath$L$}}
\renewcommand{\SS}{\mbox{\boldmath$S$}}
\newcommand{\Tc}{$T_{\rm c}$ }
\newcommand{\Tcf}{$T_{\rm c}$}
\newcommand{\Hc}{$H_{\rm c2}^{\rm P}$ }
\newcommand{\Hcf}{$H_{\rm c2}^{\rm P}$}
\newcommand{\etal}{{\it et al.}: }
\newcommand{\PRL}{Phys. Rev. Lett. } 
\newcommand{\PRB}{Phys. Rev. B } 
\newcommand{\JPSJ}{J. Phys. Soc. Jpn. } 
\newcommand{\vr}{\vec{r}}
\newcommand{\vrr}{\vec{r'}}
\newcommand{\vrrr}{\vec{r}_{1}}
\newcommand{\vrrrr}{\vec{r}_{2}}
\newcommand{\Ui}{$U_{\rm imp}$ }
\newcommand{\Uif}{$U_{\rm imp}$}

\section{Introduction}

 Superconductivity in the vicinity of quantum phase transitions  
has attracted much interest. 
 Superconductivity in high-\Tc cuprates,~\cite{Bednorz} 
organic superconductors,~\cite{Kanoda} and heavy fermions~\cite{Kitaoka} 
occurs in proximity to various quantum critical points, 
which has been a central issue in the condensed matter physics. 
 Another classical issue is the SIT, which arises from the competition 
between Anderson localization~\cite{Anderson} 
and $s$-wave superconductivity. 
 Vast theoretical studies have been devoted to the SIT triggered 
by disorders.~\cite{Maekawa,Kuroda,Finkelstein,Fisher,Ikeda,Ghosal,Scalettar,
Feigelman,Shklovskii2007,Dubi,Beloborodov,Vinokur}  
 In this paper, we point out that the recently discovered superconductivity 
in B-doped diamond~\cite{Ekimov} and 
related materials~\cite{Ren,Muranaka,Bustarret} provides a 
new field of superconductivity in the vicinity of Anderson localization.

 Diamond is recognized as a precious gem, but becomes a wide-gap 
semiconductor with light doping with substitutional boron acceptors. 
 Superconductivity has been discovered by Ekimov 
{\it et al.} by increasing the concentration of boron acceptors.~\cite{Ekimov} 
 A transition temperature higher than $10$ K has been realized 
in heavily B-doped diamond.~\cite{Takano,Umezawa} 
 It seems to be surprising that such a high transition temperature 
is induced by low carriers in a three-dimensional system
since a small density of states (DOS) near the band edge 
is harmful for superconductivity. 
 It is highly desired to identify a novel mechanism of the enhancement 
of superconductivity in B-doped diamond.

 Semiconductors have been a central field in experimental studies of 
Anderson localization.~\cite{Anderson-ex} 
 Low carriers in semiconductors are significantly affected by randomness 
since the electron wave function near the band edge is localized 
by a weak  disorder.~\cite{Belitz} 
 The effects of randomness have been actually revealed 
by experiments on B-doped diamond, which have shown 
the localization effect,~\cite{Ishizaka} 
high residual resistivity,~\cite{Sidorov,Ishizaka} 
and absence of a well-defined Fermi surface.~\cite{Yokoya} 
 Superconductivity in B-doped diamond is referred  as 
``superconductivity with no Fermi surface''.~\cite{Fukuyama}  
 However, the localization effect has never been investigated 
in theories of the B-doped diamond.~\cite{Boeri,Lee,Blase,Xiang,
Baskaran,Jafari,Shirakawa}  
 The purpose of this study is to investigate the relationship between 
localization and superconductivity in doped semiconductors.

 Two theoretical scenarios have been proposed for superconductivity 
in B-doped diamond. 
 The first one is based on the impurity band formed by impurity states 
whose wave functions are localized around boron 
atoms.~\cite{Baskaran,Jafari,Shirakawa,Nishida} 
 The other is based on the host band of diamond in which the carriers are 
provided by substitutional boron acceptors.~\cite{Boeri,Lee,Blase,Xiang} 
 The former was proposed by Baskaran~\cite{Baskaran} 
immediately after the discovery of superconductivity. 
 The latter has been proposed on the basis of first-principles band 
calculations.~\cite{Boeri,Lee,Blase,Xiang}

 Some experimental studies have been carried out to determine 
which scenario well describes the electronic structure of 
B-doped diamond.~\cite{Nakamura,Sidorov,Yokoya,Ishizaka,Wu,Hoesch} 
 Although photoemission spectroscopy~\cite{Yokoya,Ishizaka} 
has never observed any indication of the impurity band, 
the presence of the impurity band has been claimed 
by optical measurement.~\cite{Wu} 
 We point out here that the two scenarios, namely, superconductivity in 
the impurity band and that in the host band, are not contradicting concepts, 
and can be continuously described using a microscopic model. 
 In this study, we examine the crossover from the host band to the impurity band 
on the basis of the disordered attractive Hubbard model of binary alloys. 
 We show that the crossover is accompanied by the SIT, increases in \Tcf, 
and pseudogap. 
 The crossover in the electric DOS can be described by coherent potential 
approximation (CPA), which has been applied to the analysis of 
B-doped diamond.~\cite{Shirakawa} 
 However, the spatial inhomogeneity and fluctuations of superconductivity, 
which are neglected in the CPA, play an essential role in superconductivity 
in the crossover and impurity band regimes.

\begin{table}[htbp]
 \begin{tabular}{c|c|c|c|c} \hline 
& C:B & SiC:Al & SiC:B & Si:B 
\\\hline\hline 
\Tc (K) & $ > 10$ & $\sim 1.5$ & $\sim 1.5$ & $\sim 0.35$ 
\\\hline
$H_{\rm c2}$ or $H_{\rm c}$ (Oe) & $ >  10^{5}$ &$\sim 370$  &$\sim 115$ & ?
\\\hline
type & type II & type II & type I & type I
  \end{tabular}
\caption{
Superconducting properties of the B-doped diamond, SiC, and Si 
as well as the Al-doped SiC. 
The maximum values of \Tcf, and $H_{\rm c2}$ or $H_{\rm c}$ are shown. 
More detailed comparisons have been given in ref.~40.
}
\end{table}

 Boron acceptors also induce superconductivity  
in Si~\cite{Bustarret} and SiC.~\cite{Ren} 
 Recently, superconductivity has also been realized in Al-doped SiC~\cite{Muranaka}. 
 The superconducting properties of these materials are summarized in Table.~I. 
 More detailed comparisons between B-doped diamond, SiC and Si 
have been summarized by Kriener {\it et al.}~\cite{Kriener} 
 The maximum \Tc values are $0.35$ K for B-doped Si and $1.5$ K 
for B-doped SiC. 
 A more pronounced difference from B-doped diamond is the magnetic 
response in the superconducting state. 
 B-doped diamond is a type II superconductor, 
while the others are type I superconductors. 
 The large Ginzburg-Landau (GL) parameter $\kappa \sim 18$ of B-doped 
diamond~\cite{Sidorov} should be contrasted to the small GL parameter 
$\kappa \sim 0.35$ of B-doped SiC.~\cite{Kriener} 
 The upper critical field of such diamond is 
$10^{3}$ times higher than the critical field of such SiC. 
 These differences should be related to electronic properties. 
 B-doped diamond is a ``bad metal'' without any well-defined 
Fermi surface,~\cite{Yokoya} 
while B-doped SiC is a ``good metal'' with a small residual 
resistivity.~\cite{Ren,Kriener}. 
 We will propose here a theoretical interpretation for the differences 
between these superconductors.

 The paper is organized as follows. 
 The disordered attractive Hubbard model is introduced in \S2. 
 The crossover from the impurity band to the host band is demonstrated in 
\S3. 
 We examine the Anderson transition in \S4, and 
the superconductivity is investigated in \S5. 
 We discuss B-doped diamond, SiC, and Si in \S6.  
 The similarity between the doped semiconductors and 
the high-\Tc cuprates is pointed out in \S7. 
 These results are summarized in \S8.

\section{Disordered Attractive Hubbard Model}

 For a discussion of superconductivity in heavily doped semiconductors, 
we investigate the disordered attractive Hubbard model in three dimension: 
\begin{eqnarray}
\label{eq:attractive-Hubbard-model}
&& \hspace*{-10mm}
H= -t \sum_{<i,j>,\sigma} c_{i\sigma}^{\dag}c_{j\sigma} 
+ \sum_{i} (W_{i}-\mu) n_{i}
+ U \sum_{i} n_{i,\uparrow} n_{i,\downarrow}, 
\nonumber \\  \hspace*{-15mm} &&
\end{eqnarray}
where $n_{i,\sigma}$ is the number of electrons at the site $i$ with 
the spin $\sigma$ and $n_{i} =\Sigma_{\sigma} n_{i,\sigma}$. 
 The simple cubic lattice is assumed and the symbol $<i,j>$ 
denotes the summation over the nearest neighbor sites. 
 We choose the unit of energy as $t=1$. 

 The disorder is described by the random potential $W_{i}$. 
 We assume a binary alloy in which $W_{i}=0$ at host sites 
(carbon sites in B-doped diamond) and $W_{i}=U_{\rm imp}$ at impurity sites 
(boron sites in B-doped diamond). 
 We denote impurity concentration as $n_{\rm imp}$. 
 The chemical potential $\mu$ is chosen so that 
the total electron concentration is $n=2-n_{\rm imp}$ in accordance with those in  
B-doped diamond, SiC, and Si. 
 Note that the presence of a B-H complex~\cite{Mukuda} 
hardly affects the following results, and therefore 
$n_{\rm imp}$ should be regarded as the concentration of 
isolated boron atoms. 
 We assume $n_{\rm imp}=0.02$ in \S3, \S4, and \S5.2-5.5, and discuss the 
doping dependence of superconductivity in \S5.6. 

 It is believed that the electron phonon interaction gives rise to 
superconductivity in B-doped diamond.~\cite{Boeri,Lee,Blase,Xiang} 
 We take into account the static attractive interaction $U \le 0$ 
to describe the $s$-wave pairing interaction for simplicity.

\section{Crossover from Host Band to Impurity Band}

 The crossover from the host band to the impurity band is described 
by varying the impurity potential $U_{\rm imp}$ in the model eq.~(1). 
 Figure~1 shows the DOS's for various impurity potentials. 
 Here, we assume $U=0$ for simplicity.

\begin{figure}[ht]
\begin{center}
\includegraphics[height=7cm]{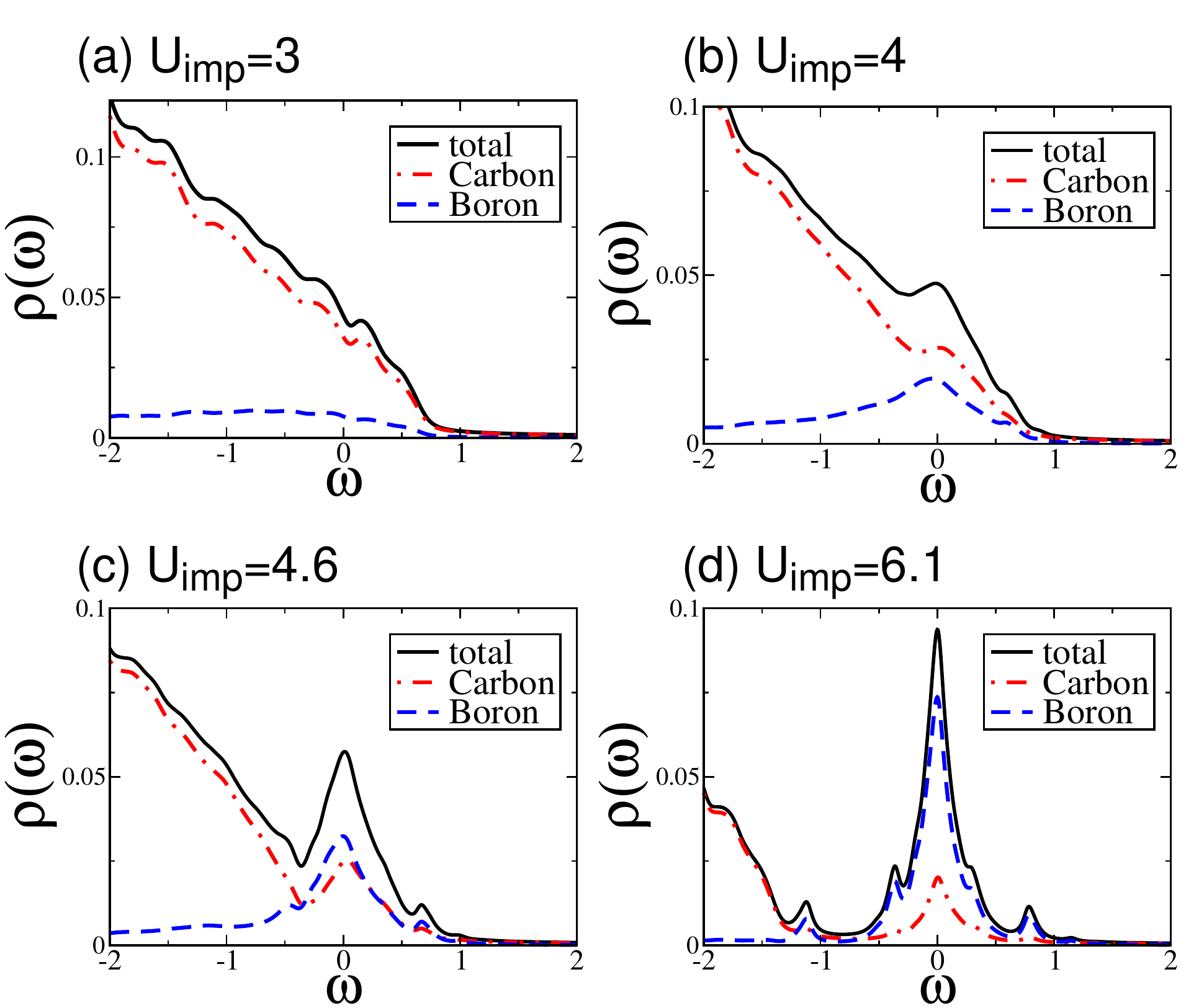}
\caption{
DOS in the normal state (solid line). 
We assume $U_{\rm imp}=3$, $4$, $4.6$, and $6.1$ in (a), (b), (c), and (d), 
respectively. 
 We assume $n_{\rm imp}=0.02$ and $U=0$. 
 The dashed and dash-dotted lines show the partial DOS's 
in the boron and carbon sites, respectively. 
 The numerical calculation is carried out in $21^{3}$ sites. 
50 samples are taken for determining the random average. 
}
\end{center}
\end{figure}

 For a large impurity potential $U_{\rm imp}=6.1$, 
the impurity band is clearly separated from the 
host band, as shown in Fig.~1(d). 
The holes are nearly half-filled in the impurity band.~\cite{Shirakawa} 
 On the other hand, the impurity band is implanted into the host band 
for a small impurity potential $U_{\rm imp}=3$, 
as shown in Fig.~1(a). 
 Single-particle states around the Fermi level are formed by impurity states 
in the former, while those are formed by the host band of diamond in the latter. 
 A crossover between the two regimes occurs, 
and the impurity band merges with the host band 
at around $U_{\rm imp}=4 \sim 4.6$, as shown in Figs.~1(b) and 1(c). 
 The acceptor level of the lightly doped boron atoms in diamond 
has been estimated to be $0.37$ eV~\cite{Ramdas}, 
which corresponds to $U_{\rm imp}=4.6$.~\cite{Shirakawa} 
 Thus, B-doped diamond seems to be near the crossover regime.

\section{Anderson Localization}

 In this section, we investigate Anderson localization 
in the model eq.~(1) to clarify the electronic structure 
in the normal state, 
which is a background of superconductivity. 
 We analyze here the inverse participation ratio (IPR) 
and level statistics for this purpose.

 For $U=0$, the Hamiltonian eq.~(1) is diagonalized 
on the basis of the single-particle states. 
 The IPR is defined as 
\begin{eqnarray}
\label{eq:IPR}
\hspace*{-10mm}
&& {\rm IPR} = 
<\sum_{\vr} |\psi_{\alpha}(\vr)|^{4}>_{\rm E_{\rm F}}, 
\end{eqnarray}
where $\psi_{\alpha}(\vr)$ is a single-particle wave function 
with an energy $\varepsilon_{\alpha}$. 
 The bracket $<>_{\rm E_{\rm F}}$ indicates the average for the 
states around the Fermi energy, $|\varepsilon_{\alpha} - \mu| < \e_{\rm c}$. 
 We choose $\e_{\rm c}=0.2$ so that the cutoff $\e_{\rm c}$ 
is less than the width of the impurity band. 
 The IPR describes the itinerant or localized character of the single-particle 
state. 
 The size scaling of the IPR shows IPR $\propto N^{-1}$ with $N$ being 
the number of sites for the completely extended state, 
while the IPR is constant for the localized state. 
 We denote here $N=L^{3}$ with $L$ being 
the linear size of the three-dimensional lattice. 
 The single particle wave function has a fractal dimension $d^{*}$ 
in disordered systems, where IPR $\propto N^{-d^{*}/d}$. 
 Note that $d=3$ in our case. 
 The universal distribution of the fractal dimension $d^{*}$ 
at the mobility edge has been shown.~\cite{Mildenberger} 
 The average is estimated to be $d^{*} = 1.3$ for the three-dimensional 
Anderson transition in the orthogonal universality class.~\cite{Mildenberger}

\begin{figure}[ht]
\begin{center}
\includegraphics[width=7cm]{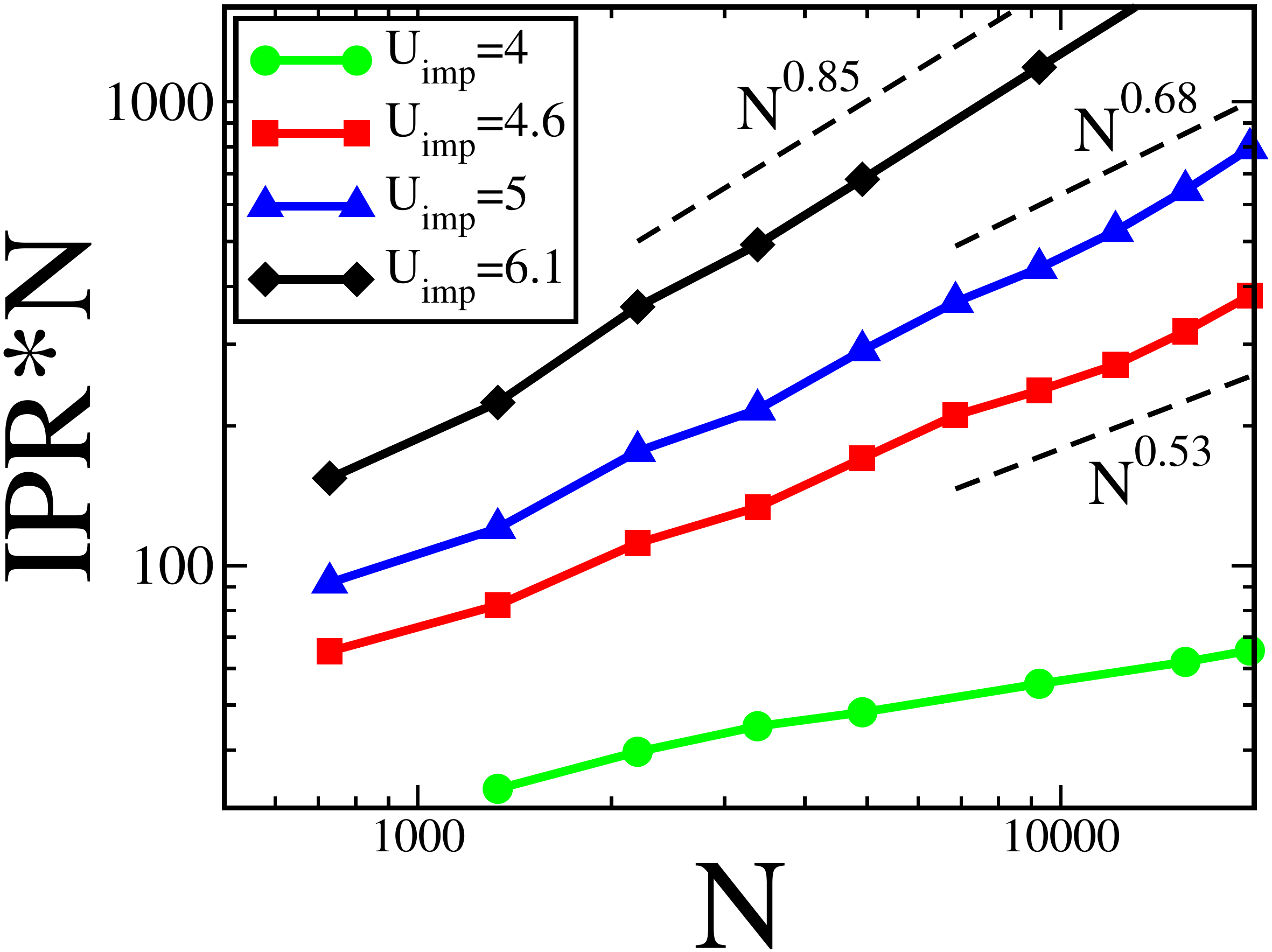}
\caption{
 Size scaling of the IPR for $U_{\rm imp}=4, 4.6, 5$, and $6.1$ 
from the bottom to the top. 
 We assume that $n_{\rm imp}=0.02$ and $U=0$. 
 The power against the size $N$ is shown as a visual guide. 
 We take into account 220, 150, 100, 70, 50, 40, 30, and 20 samples 
for $13^{3}$, $15^{3}$, $17^{3}$, $19^{3}$, $21^{3}$, $23^{3}$, $25^{3}$,  
and $27^{3}$ sites, respectively. 
}
\end{center}
\end{figure}

 Figure~2 shows the scaling plot of the IPR$\times N$ for a fixed impurity 
concentration $n_{\rm imp}=0.02$.  
 We observe a significant decrease in the fractal dimension $d^{*}$ 
with increasing impurity potential $U_{\rm imp}$ through 
the crossover from the host band to the impurity band. 
 Since the average fractal dimension $d^{*}=1.3$ at the mobility edge leads to 
the scaling IPR $\times N \propto N^{0.57}$, 
we find Anderson transition at around $U_{\rm imp}=4.6$. 
 Thus, the crossover from the host band to the impurity band 
is accompanied by Anderson localization. 
 It is reasonable that the single-particle wave function becomes localized 
with approach to the impurity band regime.

 The IPR is also an important index for the superconductivity, 
because the attractive interaction between the time-reversed states 
is proportional to the IPR. 
 Since Cooper pairing is determined by this interaction 
in the localized limit,~\cite{Feigelman} 
the data in Fig.~2 implies the enhancement of Cooper pairing 
with increasing $U_{\rm imp}$. 
 This is qualitatively true, although the parameters assumed here are 
far from the localized limit. 
 The DOS shown in Fig.~1 gives another interpretation of the enhancement 
of Cooper pairing across the crossover.  
 The effect of the attractive interaction is indicated by the ratio $|U|/W$ 
for a small \Uif, while that should be represented by $|U|/W_{\rm imp}$ 
for a large \Uif, where $W$ and $W_{\rm imp}$ are the width of the host band 
and that of the impurity band, respectively. 
 Since $W \gg W_{\rm imp}$ for a small impurity concentration $n_{\rm imp}$, 
the interaction is effectively enhanced in the impurity band region. 
 The enhanced pairing interaction leads to the enhancement of 
superconductivity in the crossover regime, as shown in \S5.

\begin{figure}[ht]
\begin{center}
\includegraphics[width=8cm]{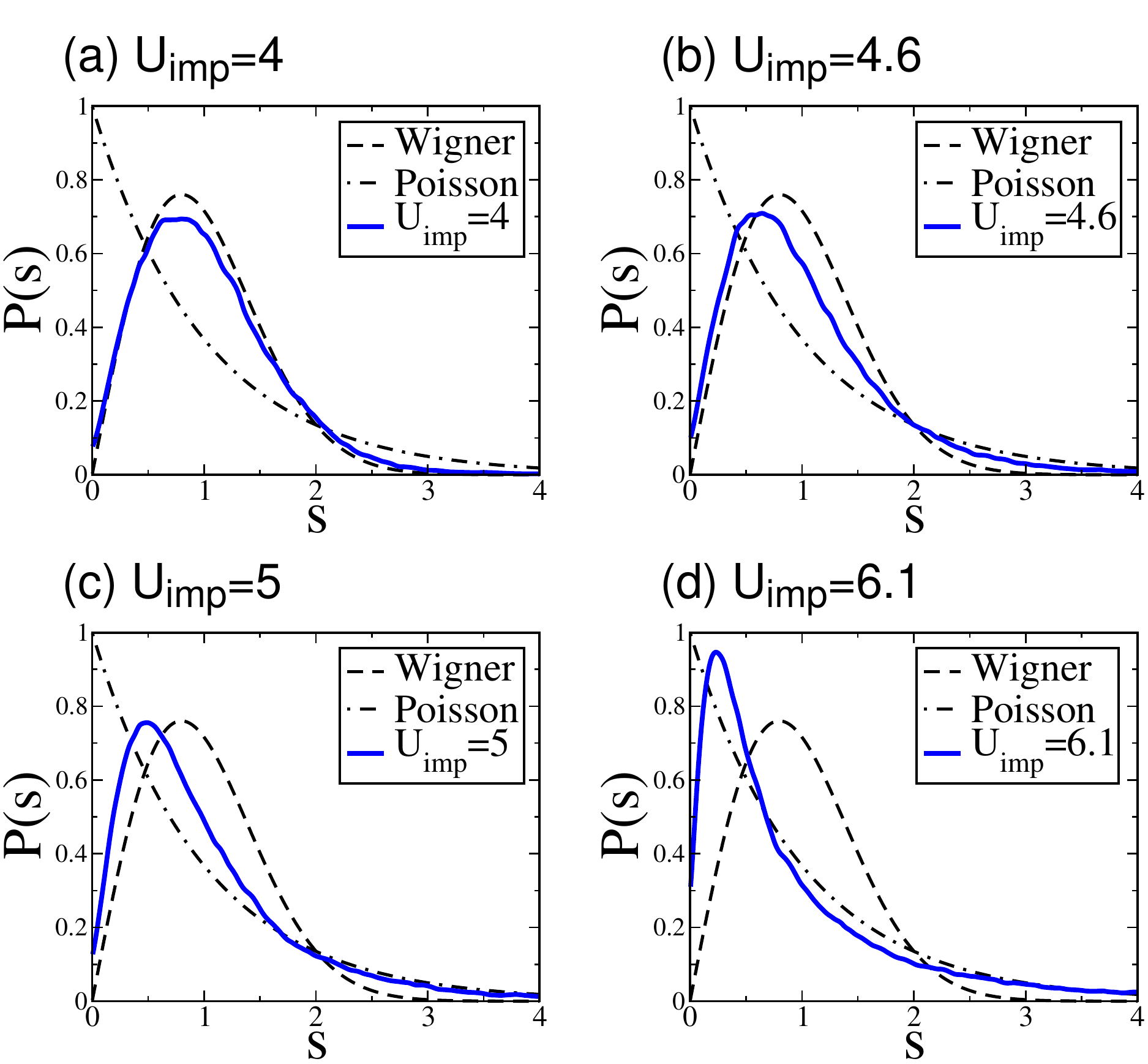}
\caption{
Results of the level statistics for $n_{\rm imp}=0.02$. 
We show the nearest-level-spacing distribution function 
$P(s)$ for energy levels within $|\e_{\alpha} - \mu| < \e_{\rm c} = 0.2$. 
The crossover from the Wigner-Dyson statistics to the Poisson statistics 
occurs as the impurity potential $U_{\rm imp}$ increases. 
The numerical calculation is carried out for $21^{3}$ sites and 250 samples. 
}
\end{center}
\end{figure}

 Anderson localization in the crossover regime is also observed 
by the analysis of level statistics.~\cite{Shklovskii-level} 
 Figure~3 shows the nearest-level-spacing distribution function $P(s)$ 
defined as 
\begin{eqnarray}
\label{eq:level-statistics}
\hspace*{-10mm}
&& P(s) = 
<\delta(|\e_{\alpha}-\e_{\alpha+1}|/\e_{\rm s} -s)>_{E_{\rm F}}, 
\end{eqnarray}
where $\e_{\rm s}$ is the averaged nearest level spacing. 
 It has been shown that the distribution obeys the Poisson statistics 
$P(s)=\exp(-s)$ in the insulating state,  
while it obeys the Wigner-Dyson statistics 
$P(s)=\frac{\pi}{2} s \exp(-\pi s^{2}/4)$ 
in the metallic state.~\cite{Wigner-Dyson,Shklovskii-level}  
 As shown in Fig.~3, the nearest-level-spacing distribution function $P(s)$ 
shows a crossover from the Wigner-Dyson statistics to the Poisson statistics 
with increasing \Ui across the crossover from the host band to the impurity band. 
 Because of the small DOS near the band edge, it is difficult to carry out 
finite-size scaling.~\cite{Shklovskii-level} 
 However, it is clear that the Anderson transition occurs 
in the crossover regime at around $U_{\rm imp}=4.6$,  
consistent with the analysis of IPR.

\section{Superconductivity}

\subsection{Formulation}

 Next, we discuss the superconductivity.  
 We adopt real-space self-consistent T-matrix approximation 
(RSTA) to analyze the model eq.~(1). 
 The RSTA has been formulated for disordered $d$-wave 
superconductivity in high-\Tc cuprates~\cite{Yanase2006}. 
 The thermal fluctuation of superconductivity 
is taken into account in the self-consistent $1$-loop order. 
 Because the calculation is carried out in real space, 
randomness is accurately taken into account. 

 Most microscopic studies of superconductivity in random systems 
have been based on mean field 
approximation.~\cite{Franz,Ghosal,Atkinson,HirschfeldBdG}
 However, the fluctuation plays an important role in strongly disordered 
systems because the fluctuation is enhanced by microscopic 
inhomogeneity.~\cite{Yanase2006,Ghosal} 
 RSTA can be carried out in more than $10^{3}$ sites, which is needed  
for the analysis of three-dimensional models. 
 We carry out a numerical calculation in a $11 \times 11 \times 11$ lattice, 
unless otherwise mentioned. 
 This lattice size is much larger than the limit of quantum~\cite{Scalettar} 
and classical~\cite{Dubi} Monte Carlo simulations.

 The formulation of RSTA is described as follows. 
 We diagonalize the non-interacting Hamiltonian   
\begin{eqnarray}
\label{eq:H_0}
&& \hspace*{-12mm} 
H_{0}= -t \sum_{<i,j>,\sigma} c_{i\sigma}^{\dag}c_{j\sigma} 
+ \sum_{i} (W_{i}-\mu) n_{i},  
\end{eqnarray}
and obtain the undressed Green function as 
\begin{eqnarray}
\label{eq:G_0}
&& \hspace*{-10mm}  
G_{0}(\vr,\vrr,\omega_{n}) = 
\sum_{\alpha} \psi_{\alpha}(\vr) 
\frac{1}{{\rm i}\omega_{n} - \varepsilon_{\alpha}}
\psi^{*}_{\alpha}(\vrr), 
\end{eqnarray}
where $\omega_{n}=(2 n +1) \pi T$ is the Matsubara frequency.

 The Dyson equation is expressed as  
\begin{eqnarray}
\label{eq:G}
&& \hspace*{-12mm}  
G(\vr,\vrr,\omega_{n})= 
G_{0}(\vr,\vrr,\omega_{n}) + 
\nonumber \\ && \hspace*{0mm}
\sum_{\vrrr,\vrrrr}
G_{0}(\vr,\vrrr,\omega_{n})
\Sigma(\vrrr,\vrrrr,\omega_{n})
G(\vrrrr,\vrr,\omega_{n}), 
\end{eqnarray}
where $\Sigma(\vrrr,\vrrrr,\omega_{n})$ is the self-energy 
represented in real space. 
 We estimate self-energy using the 
self-consistent 1-loop-order approximation,  
whose results in the clean limit have been summarized 
in ref.~50. 
 We adopt quasi-static 
approximation~\cite{Lee-Rice-Anderson,Sadovskii,Tchernyshyov,SchmalianPG,
Yanase2004} as in ref.~46, where quantum fluctuation is ignored, 
but thermal fluctuation is appropriately taken into account. 
 We have quantitatively shown the validity of quasi-static approximation 
as well as self-consistent 1-loop-order approximation 
for superconducting fluctuation.~\cite{Yanase2004} 
 Note that the characteristic dynamical property of superconducting 
fluctuation leads to the validity of quasi-static 
approximation.~\cite{Yanase2004}

 Self-energy is given as 
\begin{eqnarray}
\label{eq:self-energy}
&& \hspace*{-10mm} \Sigma(\vr,\vrr,\omega_{n}) = 
- T U^{2}  T(\vr,\vrr) G(\vrr,\vr,-\omega_{n}).  
\end{eqnarray}
 The propagator of superconducting fluctuation, 
namely, the T-matrix, is obtained using 
\begin{eqnarray}
\label{eq:T-matrix}
&& \hspace*{-12mm} 
T(\vr,\vrr) = T_{0}(\vr,\vrr) - 
\sum_{\vrrr}
U T_{0}(\vr,\vrrr)  
T(\vrrr,\vrr), 
\\
\label{eq:bare-T-matrix}
&&  \hspace*{-12mm} 
T_{0}(\vr,\vrr)
=T \sum_{n} G(\vr,\vrr,\omega_{n}) G(\vr,\vrr,-\omega_{n}) \phi(\omega_{n})^{2}, 
\end{eqnarray} 
where we take into account the cutoff frequency as 
$\phi(\omega) = [1+\exp\{a(|\omega|/\omega_{c}-1)\}]^{-1}$ 
with $a=20$.  
 We assume the cutoff frequency $\omega_{\rm c}=0.2$ 
in accordance with the phonon frequency $\omega_{\rm ph} \sim 150$ meV 
in B-doped diamond.~\cite{Hoesch} 
 The T-matrix diverges at the superconducting transition temperature. 
 This criterion gives rise to the transition temperature of the BCS theory, 
namely, the mean field theory with respect to $U$, 
when we neglect self-energy.

 Since the random potential is included in the unperturbed Hamiltonian $H_{0}$, 
randomness is accurately taken into account in  RSTA. 
 The spatial dependences of superconducting fluctuation and 
single-particle states are self-consistently determined. 
 We find that perturbative approximations for randomness, 
such as the Born approximation,~\cite{Abrikosov} CPA,~\cite{Shirakawa} 
and T-matrix approximation,~\cite{Maki,Hotta,Hirschfeld} break down 
for a large impurity potential $U_{\rm imp} > 4$. 
 This is because the quasiparticle states around the Fermi level 
mainly consist of impurity states whose wave function is localized 
around impurity sites. 
 Impurity potential should be taken into account in zeroth-order 
approximation when the impurity band plays an essential role in 
superconductivity.

\subsection{Microscopic inhomogeneity}

 First, we investigate the spatial inhomogeneity 
of superconducting susceptibility defined as 
\begin{eqnarray}
\label{eq:chi-space}
&& \hspace*{-12mm} 
\chi_{\rm sc}(\vr) = \sum_{\vrr} T(\vr,\vrr), 
\end{eqnarray}
which is spatially uniform in the clean limit.  
 A typical spatial dependence of $\chi_{\rm sc}(\vr)$ is shown in Fig.~4 
for various \Ui values. 
 Since it is difficult to show the spatial dependence in the entire 
three-dimensional lattice, $\vr=(x,y,z)$, 
we show the spatial dependence on the plane $z=1$. 

\begin{figure}[ht]
\begin{center}
\vspace*{-5mm}
\hspace*{-20mm}
\includegraphics[width=12cm]{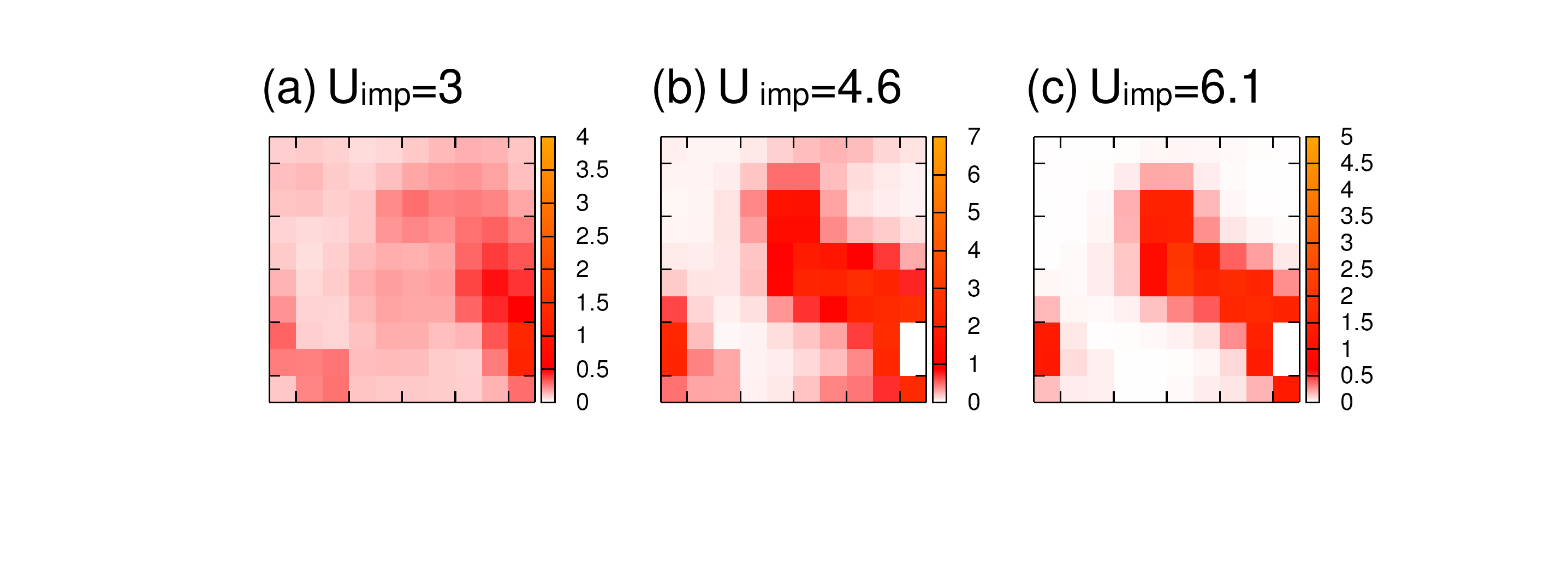}
\caption{Typical spatial dependences of 
the superconducting susceptibility $\chi_{\rm sc}(\r)$ 
for (a) $U_{\rm imp}=3$, (b) $U_{\rm imp}=4.6$, and (c) $U_{\rm imp}=6.1$. 
We assume $n_{\rm imp}=0.02$, $T=0.002$, and $U=-1$. 
We show the results on the plane at $z=1$ in the three-dimensional lattice. 
}
\end{center}
\end{figure}

 In Figs.~4(a)-4(c), the superconducting correlation develops in the 
``dirty region'' that includes many impurities, in contrast to 
the conventional spatial dependence, in which superconductivity is 
suppressed by the impurities~\cite{Franz,Ghosal,Atkinson,HirschfeldBdG,Yanase2006}. 
 This is because the wave functions of quasiparticle states around the Fermi level 
have a large weight around impurity sites, and therefore, 
impurity sites play more important roles in superconductivity 
than host sites regardless of $U_{\rm imp} > 0$. 
 Thus, a unique spatial structure of superconductivity 
is realized in heavily doped semiconductors. 
 The $s$-wave symmetry of superconductivity is essential  for this result, 
as discussed in \S7.

 The spatial inhomogeneity of the superconducting correlation, 
namely, the mesoscopic fluctuation~\cite{Mesoscopic} increases  
with increasing $U_{\rm imp}$. 
 Superconductivity is spatially homogeneous 
in the host band regime $U_{\rm imp} < 4$, 
but significantly inhomogeneous 
in the impurity band regime $U_{\rm imp} > 5$, as shown in Fig.~4. 
 The microscopic inhomogeneity of the superconducting correlation 
can be regarded as the localization of Cooper pairs.~\cite{Feigelman} 
 Figure~4 indicates that the Anderson localization of single-particle states 
is accompanied by the localization of Cooper pairs. 
 A similar microscopic inhomogeneity has been discussed for high-\Tc cuprates 
from both experimental~\cite{Pan} and theoretical points of 
view~\cite{Dubi,Yanase2006,Franz,Atkinson,Yanase2007,HirschfeldBdG,Meyr}.

\subsection{Superconducting fluctuation and pseudogap}

 The spatial inhomogeneity of the order parameter 
generally disturbs the long-range coherence of superconductivity. 
 Then, the short-range correlation induces a pseudogap 
in the excitation spectrum. 
 In other words, the fluctuation is enhanced by the localization of 
Cooper pairs, and therefore the precursor of superconductivity 
appears above \Tcf.

 Figure~5 shows the DOS above \Tcf, which is obtained using  
\begin{eqnarray}
\label{eq:average-DOS}
&& \hspace*{-12mm} 
\rho(\omega) = <- \frac{1}{\pi N} 
\sum_{\vr} {\rm Im}G^{\rm R}(\vr,\vr,\omega)>_{\rm av}, 
\end{eqnarray}
where $<>_{\rm av}$ denotes the random average. 
 We take the random average for 10 samples of the impurity distribution. 
 We find that the DOS is self-averaged, and therefore the 
sample dependence is weak for the spatially averaged DOS. 
 We see that the effect of superconducting fluctuation is negligible 
in the host band regime ($U_{\rm imp}=4$), 
while the pseudogap is induced by a short-range 
superconducting fluctuation in the crossover and impurity regimes 
($U_{\rm imp}=4.6, 5,$ and $6.1$). 
 It is shown that the pseudogap increases with  
increasing impurity potential, and therefore 
the DOS at the Fermi energy is significantly suppressed 
in the impurity band regime above \Tcf. 

\begin{figure}[ht]
\begin{center}
\includegraphics[width=8cm]{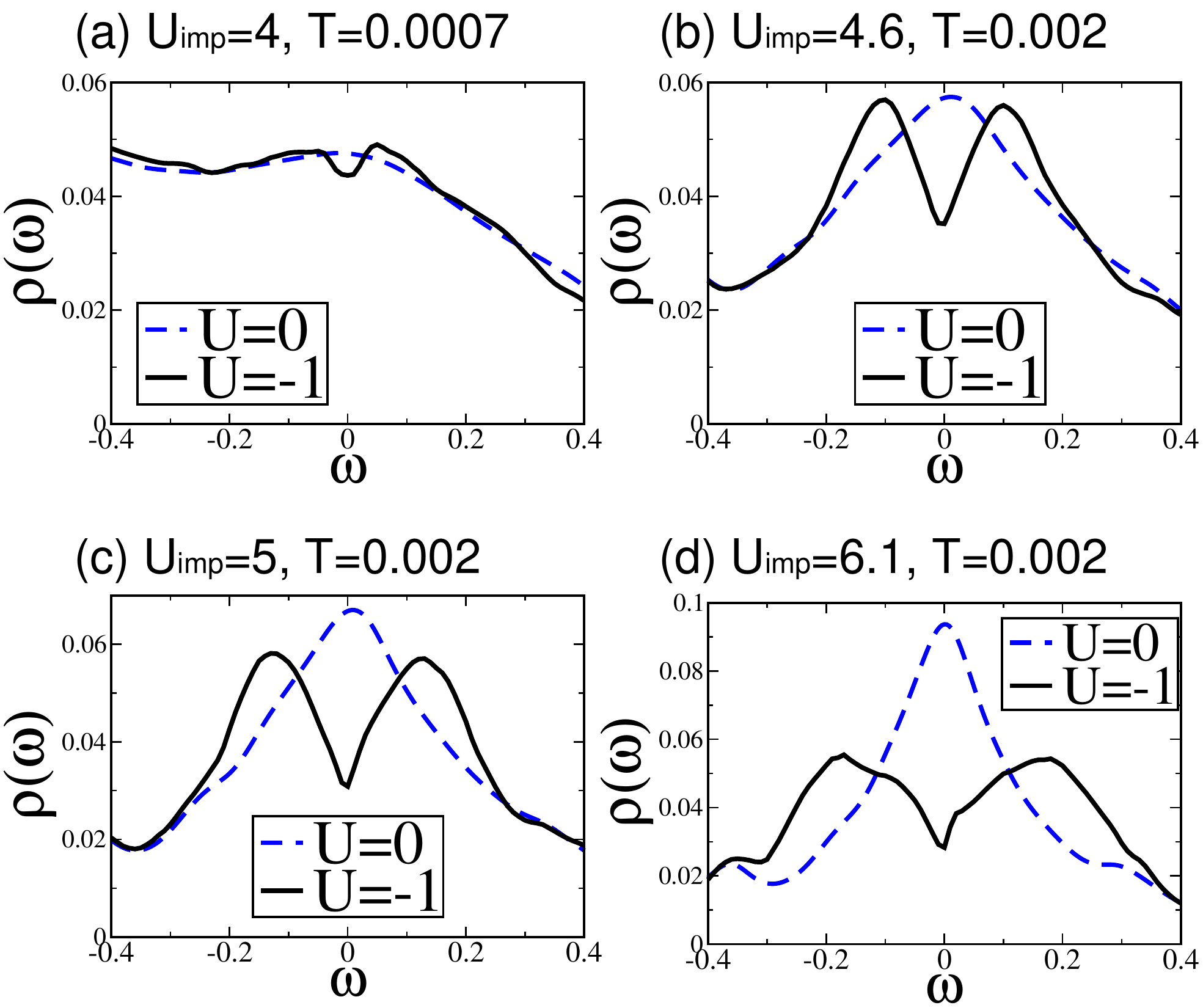}
\caption{
 DOS's for various impurity potentials $U_{\rm imp}$ (solid lines). 
 We choose the temperature close to the critical point of 
the superconductivity, 
$T=0.0007$ for $U_{\rm imp}=4$, 
and $T=0.002$ for $U_{\rm imp}=4.6, 5,$ and $6.1$.  
 We take the random average over 10 samples of the impurity distribution. 
 We assume $n_{\rm imp}=0.02$ and $U=-1$. 
 The bare DOS at $U=0$ is shown for a comparison (dashed lines). 
}
\end{center}
\end{figure}

 We explain here the mechanism of pseudogap in detail. 
 (1) The small width of the impurity band leads to a small Fermi energy,  
and therefore a small coherence length. 
 A pseudogap generally appears in small-coherence-length superconductors, 
such as high-\Tc cuprates and organic superconductors~\cite{YanaseReview}. 
 (2) The fractal structure of the single-particle wave function enhances 
the pseudogap, as discussed by Feigel'man {\it et al.}~\cite{Feigelman} 
 (3) The localization of Cooper pairs, namely, the microscopic inhomogeneity 
of superconductivity enhances the thermal fluctuation 
and also the pseudogap.~\cite{Yanase2006} 
 (4) The effective pairing interaction for the $s$-wave superconductivity 
is enhanced by the localization of single particle-states, 
as discussed in \S4. 
 The transition temperature in the mean field theory $T_{\rm c}^{\rm MF}$ 
with respect to $U$, where the fluctuation is neglected,  
increases with increasing impurity potential \Uif. 
 These features give rise to the pseudogap in the crossover 
and impurity band regimes.

 The pseudogap disappears in the host band regime because of the 
wide host band of diamond, the extended wave function of quasiparticle states,  
and the spatially homogeneous superconducting correlation. 
 Thus, the crossover from host band superconductivity to  
impurity band superconductivity is accompanied by the crossover from 
the conventional Fermi liquid state to the pseudogap state.

 A large attractive interaction $U=-9.2$ gives rise to the transition 
temperature $T_{\rm c} \sim 0.002$ in the clean limit ($U_{\rm imp}=0$). 
 We confirmed that the superconducting fluctuation is negligible for these 
parameters. The thermal fluctuation at $T = 0.002$ hardly affects the DOS 
in the absence of disorders. 
 This means that randomness plays an essential role 
in the pseudogap in the crossover and impurity band regimes.

\subsection{Superconducting susceptibility}

 We investigate here the superconducting susceptibility averaged over real space 
as well as the randomness 
\begin{eqnarray}
\label{eq:average-susceptibility}
&& \hspace*{-12mm} 
\chi_{\rm sc} = <\frac{1}{N} \sum_{\vr} \chi_{\rm sc}(\vr)>_{\rm av}. 
\end{eqnarray}
 This susceptibility diverges at the critical point of superconductivity. 
 Although the superconducting susceptibility is always finite in our calculation 
due to the finite size effect, the magnitude of $\chi_{\rm sc}$ 
indicates the closeness to the critical point.

\begin{figure}[ht]
\begin{center}
\includegraphics[width=6.5cm]{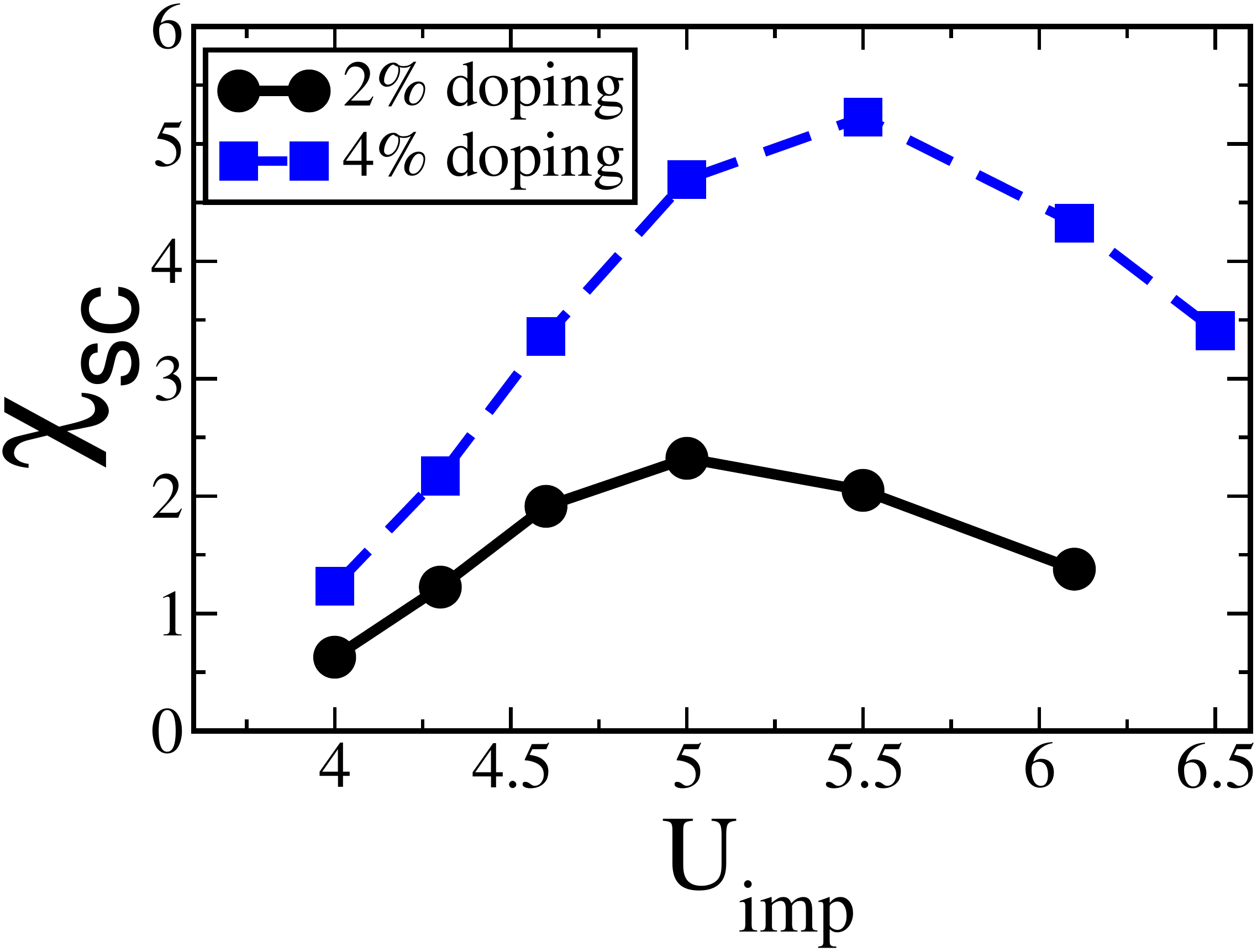}
\caption{
Superconducting susceptibility $\chi_{\rm sc}$. 
We show the \Ui dependences for $n_{\rm imp}=0.02$ (circles) and 
$n_{\rm imp}=0.04$ (squares), respectively. 
 We assume $T=0.002$ and $U=-1$. 
}
\end{center}
\end{figure}

 Figure~6 shows the \Ui dependences of $\chi_{\rm sc}$ 
for $n_{\rm imp}=0.02$ and $n_{\rm imp}=0.04$. 
 We see the dome shape of superconducting susceptibility 
against the impurity potential for a fixed impurity concentration. 
 The peak appears in the crossover regime. 
 These results indicate that the optimum \Tc is obtained  
in the crossover regime.

\begin{figure}[ht]
\begin{center}
\includegraphics[width=6.5cm]{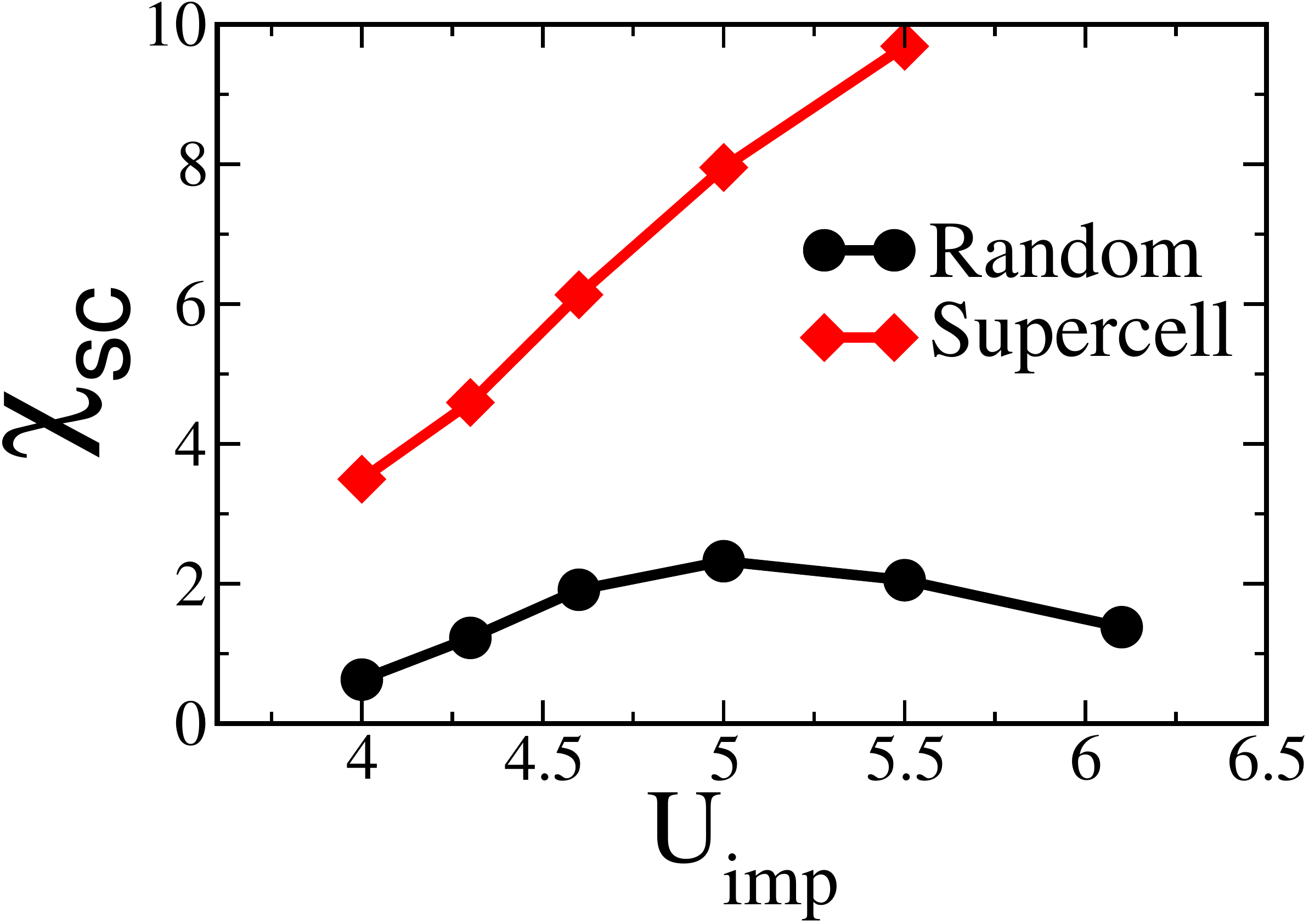}
\caption{
 Superconducting susceptibility $\chi_{\rm sc}$ 
estimated for the supercell (diamonds). 
 We put the impurities at $\vr = (3 l +1, 3 m +1, 3 n +1)$ 
with $l,$ $m,$ and $n$ being integers. 
The impurity concentration is $n_{\rm imp}=0.0203$ for the 
$11 \times 11 \times 11$ lattice. 
 The $\chi_{\rm sc}$ in the random system with $n_{\rm imp}=0.02$ 
is shown for comparison (circles). 
 We assume $T=0.002$ and $U=-1$.
}
\end{center}
\end{figure}

 The randomness is essential for the dope shape of the superconducting 
susceptibility in Fig.~6. 
 To illuminate this point, we show the superconducting susceptibility 
$\chi_{\rm sc}$ in the supercell, 
in which impurities are set at $\vr = (3 l +1, 3 m +1, 3 n +1)$. 
 Figure~7 shows the comparison between the supercell 
and the random system. 
 We see that the susceptibility $\chi_{\rm sc}$ monotonically increases 
in the supercell with increasing \Uif, in contrast to the random system. 
 Thus, the suppression of the superconductivity in the impurity band regime 
rather arises from randomness.

 The enhancement of superconductivity in the supercell has been 
pointed out by Shirakawa {\it et al.}~\cite{Shirakawa} on the basis of the CPA. 
 Our results are consistent with their claim. 
 However, superconductivity is enhanced in the supercell in our 
calculation because fluctuation, which is not taken into account 
in the CPA, is suppressed.

 We have confirmed that the transition temperature in the mean field theory, 
$T_{\rm c}^{\rm MF}$, monotonically increases with increasing  
impurity potential \Ui owing to the enhancement of pairing interaction, 
as discussed in \S4. 
 On the other hand, superconducting susceptibility is suppressed 
by mesoscopic and thermal fluctuations in the crossover and 
impurity band regimes. 
 Thus, the competition between the enhancement of Cooper pairing and that 
of superconducting fluctuations gives rise to the dome shape of 
superconducting susceptibility in Fig.~6. 

 The long-range coherence of superconductivity is hardly realized 
in the impurity band regime because of the localization of 
Cooper pairs (see \S5.2). 
 The susceptibility defined in eqs.~(10) and (12)  
takes into account both the short-range and long-range correlation 
of superconductivity. 
 The weight of the short-range and long-range correlation can be 
clarified by analyzing the averaged correlation function  
\begin{eqnarray}
\label{eq:average-correlation}
&& \hspace*{-12mm} 
T_{\rm av}(\vr) = <\frac{1}{N} \sum_{\vrr} T(\vrr+\vr,\vrr)>_{\rm av}. 
\end{eqnarray}
 We have confirmed that the correlation function increases
for a small $|\vr|$ in the impurity band regime, but then decreases 
for a large $|\vr|$ with increasing \Uif. 
 This means that the short-range correlation develops, 
but the long-range correlation is suppressed in the impurity band regime, 
as we have mentioned above.

 Figure~6 shows that the peak of $\chi_{\rm sc}$ shifts to 
the impurity band regime with increasing impurity concentration $n_{\rm imp}$. 
 This is because Anderson localization is suppressed by increasing 
the concentration of carriers, $n_{\rm imp}$, 
and then superconducting fluctuation is also suppressed. 
 The pseudogap is suppressed by increasing $n_{\rm imp}$,  
while the \Tc is increased, as will be shown in Figs.~9-11. 

\subsection{Phase diagram}

 We illustrate here a schematic figure of the phase diagram 
against the impurity potential $U_{\rm imp}$ 
and the temperature $T$ (Fig.~8).

\begin{figure}[ht]
\begin{center}
\includegraphics[width=7cm]{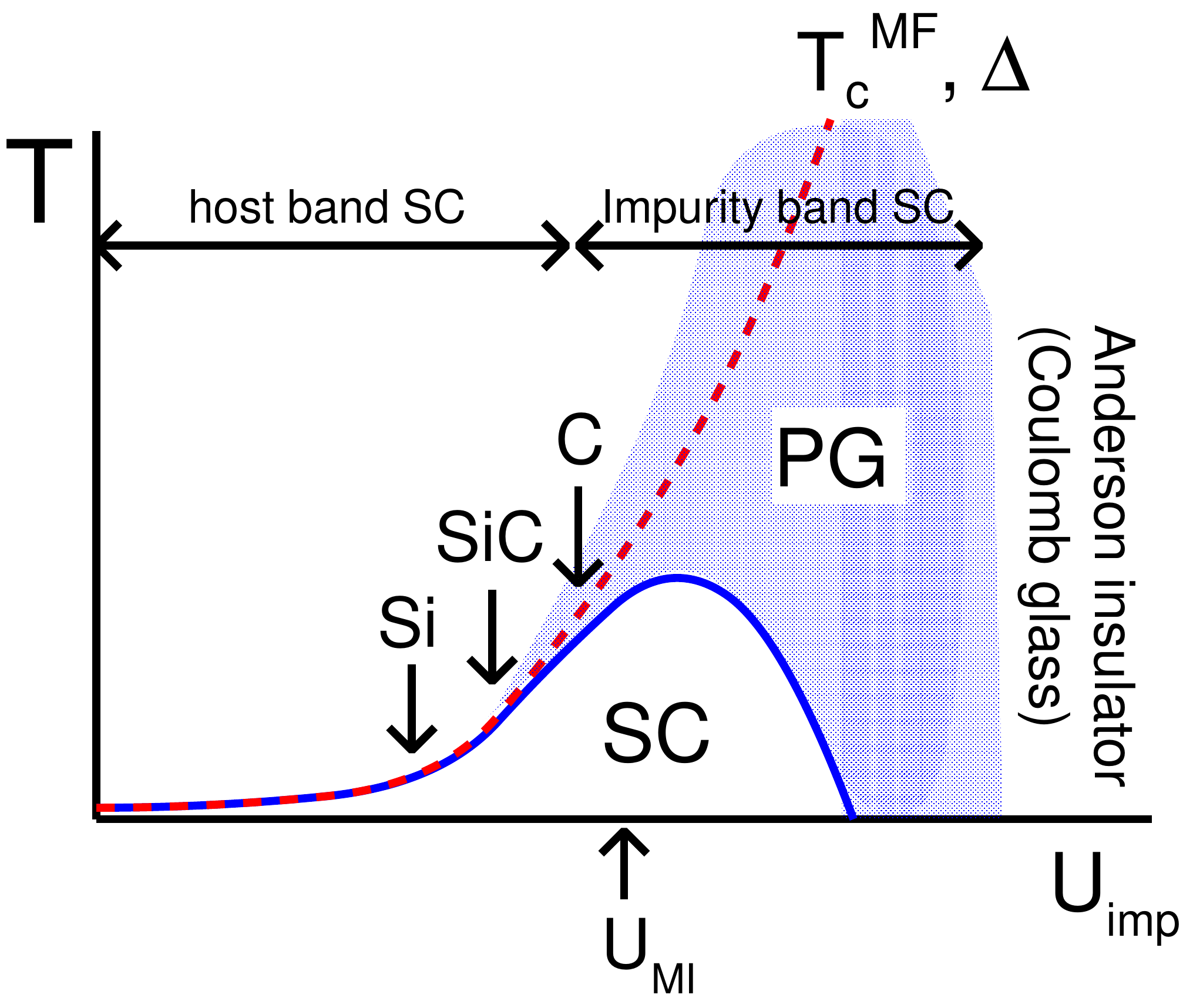}
\caption{
Schematic figure of the phase diagram against the impurity potential 
$U_{\rm imp}$ and the temperature $T$. 
The crossover from the host band to the impurity band is 
described. 
The solid line shows the transition temperature \Tc of superconductivity. 
The dashed line shows the transition temperature $T_{\rm c}^{\rm MF}$ 
in the mean field theory. 
The shaded region indicates the pseudogap state induced by 
the incoherent Cooper pairs. 
The upward arrow ($U_{\rm MI}$) shows the virtual quantum metal-insulator 
transition in the absence of superconductivity. 
The downward arrows indicate our interpretation for B-doped diamond, SiC, 
and Si (see \S6). 
}
\end{center}
\end{figure}

 As we have discussed in \S4, the pairing interaction is effectively enhanced 
by the localization of single-particle wave functions with increasing \Uif. 
 Therefore, the transition temperature $T_{\rm c}^{\rm MF}$ 
in the mean field theory and the magnitude of superconducting gap $\Delta$ 
at $T=0$ monotonically increase with increasing impurity potential 
\Ui (dashed line in Fig.~8). 
 We have confirmed this \Ui dependence of the superconducting gap  
in the mean field theory at $T=0$.

 The Anderson localization of single-particle wave functions
leads to microscopic inhomogeneity, which enhances  
the thermal fluctuation in superconductivity. 
 Then, the long-range coherence is disturbed, as shown in \S5.4. 
 The competition between the enhancement of the effective pairing interaction 
and that of superconducting fluctuations leads to a dome shape of 
the superconducting state with a peak \Tc in the crossover regime 
(solid line in Fig.~8).

 The short-range correlation of superconductivity gives rise to the 
resonance scattering between incoherent Cooper pairs and quasiparticles, 
which leads to a pseudogap in the shaded region in Fig.~8. 
 In other words, the superconducting gap obtained in the mean field theory 
is changed to a pseudogap by the fluctuations. 

 The pseudogap state would be an insulating state 
in the highly disordered regime because both electrons 
and Cooper pairs are localized. 
 This state is regarded as the ``hard gap insulator'' 
proposed by Feigel'man {\it et al.}~\cite{Feigelman}, 
which seems to be realized in the disordered thin films.~\cite{Haviland,Hebard,
Gantmakher,Sambandamurthy,Crane,Baturina,Spathis,SacepeTiN,Sarwa,
SteinerInO,Parendo,Stewart} 
 Thus, the insulating state near the SIT is expected to be 
the localized Cooper pairing state, but not the Coulomb glass state.~\cite{Efros}

\subsection{Doping dependence}

 We investigate here the doping dependence. 
 Impurity level depends on the impurity concentration $n_{\rm imp}$ 
in the heavily doped semiconductors, but we show the results for 
a fixed $U_{\rm imp}$ for simplicity.

\begin{figure}[ht]
\begin{center}
\includegraphics[width=6cm]{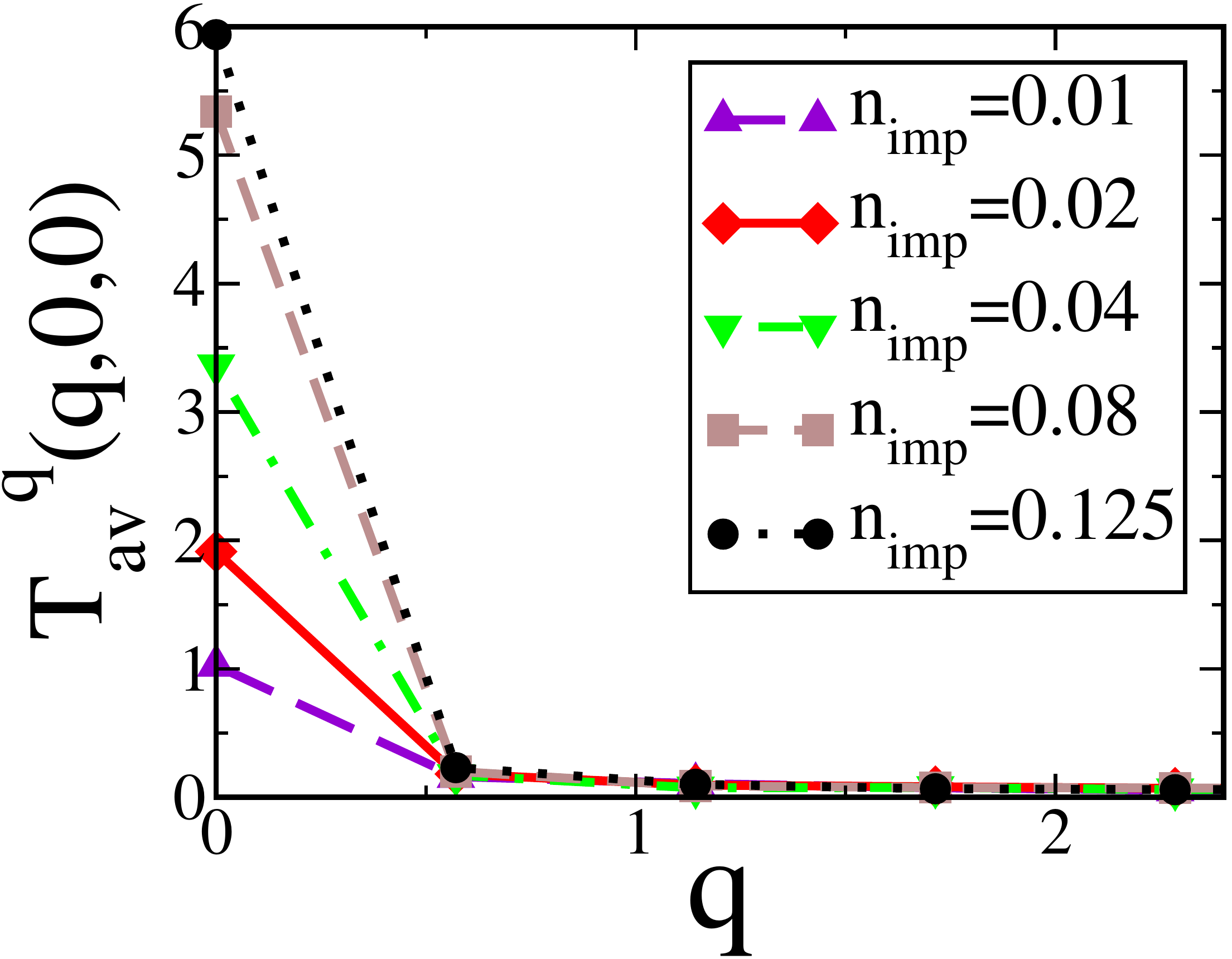}
\caption{Fourier-transformed correlation function 
$T_{\rm av}^{\rm q}(\q)$ along $\q = q \hat{x}$ for 
$n_{\rm imp}=0.01$, $0.02$, $0.04$, $0.08$, and $0.125$ from the bottom 
to the top. We assume $U_{\rm imp}=4.6$, $U=-1$, and $T=0.002$. 
}
\end{center}
\end{figure}

 Figure~9 shows the doping dependence of the 
Fourier-transformed correlation function 
\begin{eqnarray}
\label{eq:Fourier-correlation}
&& \hspace*{-12mm} 
T_{\rm av}^{\rm q}(\q) = \int T_{\rm av}(\vr) {\rm e}^{{\rm i} \q \r} {\rm d}\q. 
\end{eqnarray}
 The spatially averaged superconducting susceptibility $\chi_{\rm sc}$ is related to 
this correlation function as $\chi_{\rm sc} = T_{\rm av}^{\rm q}(0)$. 
 We see that the correlation function increases at $\q=0$ with increasing  
doping concentration; it hardly changes at $\q \ne 0$. 
 This means that the long-range correlation of superconductivity 
develops in the heavily doped regime. 
 This is consistent with the experimental observation that 
the \Tc of B-doped diamond increases with increasing 
concentration of isolated boron acceptors.~\cite{Takano,Umezawa,Mukuda}

 It has been pointed out that the concentration of carriers is 
different from that of boron atoms owing to the presence of 
the B-H complex and/or the B-B dimer.~\cite{Mukuda,Nakamura2}  
 We have confirmed that the B-H complex as well as the B-B dimer neither 
enhance nor suppress superconductivity.
 Therefore, the doping concentration $n_{\rm imp}$ discussed in this paper 
should be regarded as the concentration of isolated boron acceptors, 
consistent with the experimental observation.~\cite{Mukuda}

\begin{figure}[ht]
\begin{center}
\includegraphics[width=8cm]{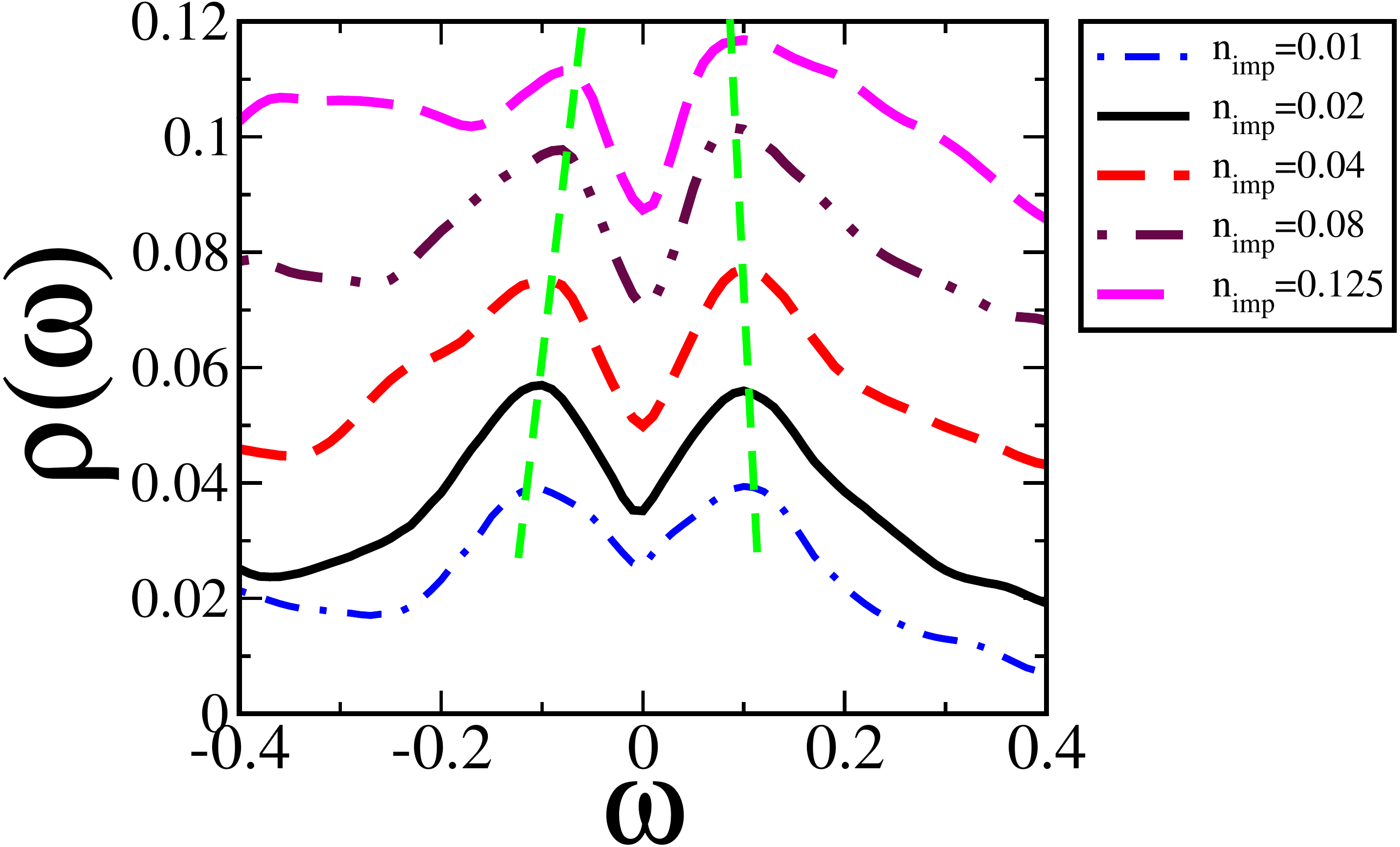}
\caption{DOS's for 
$n_{\rm imp}=0.01$, $0.02$, $0.04$, $0.08$, and $0.125$ 
from the bottom to the top. 
We assume that $U_{\rm imp}=4.6$, $U=-1$, and $T=0.002$.  
 The vertical dashed line is drawn to emphasize 
the doping dependence of the gap in DOS. 
}
\end{center}
\end{figure}

 Figure~10 shows the doping dependence of the DOS in the crossover regime 
($U_{\rm imp}=4.6$). 
  We see that the pseudogap is suppressed by increasing the doping concentration 
$n_{\rm imp}$, although the superconducting correlation is enhanced, 
as shown in Fig.~9. 
 The single-particle states around the Fermi level become itinerant 
with increasing $n_{\rm imp}$ owing to the percolation of impurity states. 
 Then, Anderson localization is suppressed, and therefore 
superconducting fluctuation is suppressed. 

\begin{figure}[ht]
\begin{center}
\includegraphics[width=7.5cm]{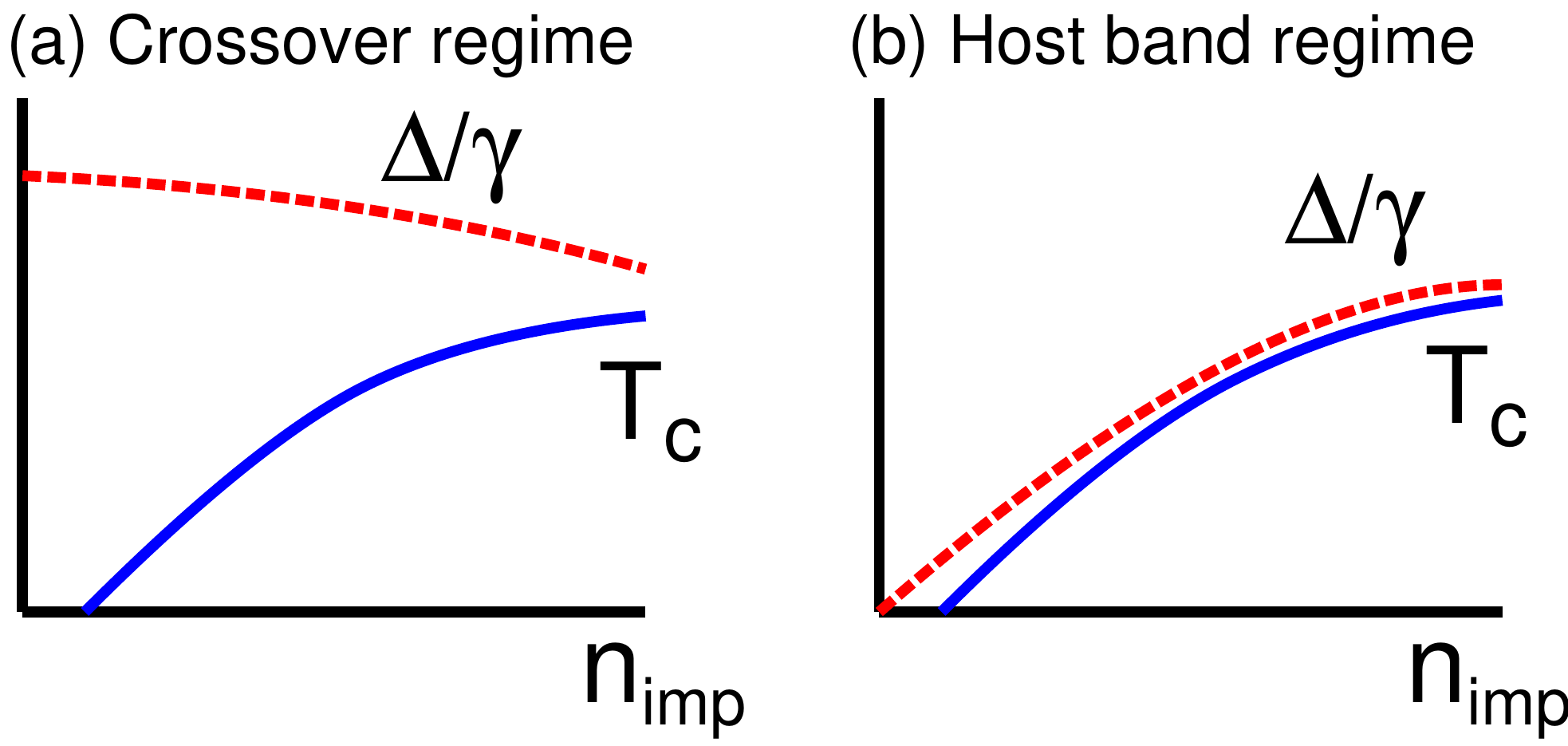}
\caption{Schematic figure of the doping dependence in (a)  
the crossover regime and (b) the host band regime, respectively. 
 We show the magnitude of the superconducting gap (or pseudogap) 
divided by $\gamma = 1.73$ for a comparison with \Tcf. 
}
\end{center}
\end{figure}

 The schematic figure of the doping dependence in the crossover regime 
is shown in Fig.~11(a). 
 We would like to stress again that the transition temperature \Tc 
of superconductivity increases, 
but the thermal fluctuation is suppressed by acceptor doping. 
 The pseudogap and superconducting gap decrease 
in contrast to \Tc with increasing doping concentration. 
 These features in the crossover regime should be contrasted to 
those in the host band regime where the thermal fluctuation is 
negligible except at a very low concentration of carriers. 
 A conventional doping dependence is expected in the host band regime;  
superconducting gap increases together with \Tcf, 
as shown in Fig.~11(b).

 We have investigated the effect of doping compensation due to doped donors 
and carrier increase due to the additional acceptors. 
 We found that superconductivity is significantly 
suppressed.~\cite{YanaseIWSDRM2008}  
 In other words, high \Tc of superconductivity is expected 
in uncompensated semiconductors with large carrier concentration.

\section{Interpretation for B-doped Diamond, SiC and Si}

 We discuss here B-doped diamond, SiC, and Si on the basis of the 
results in \S5. 
 Since no clear impurity band is observed in the angle-resolved 
photoemission spectroscopy (ARPES),~\cite{Yokoya} 
B-doped diamond does not seem to be in the impurity band regime. 
 On the other hand, the high transition temperature 
$T_{\rm c} >10$ K,~\cite{Takano,Umezawa} 
high upper critical field $H_{\rm c2} > 10$ T,~\cite{Takano,Umezawa,Sidorov} 
and large Ginzburg-Landau (GL) parameter $\kappa \sim 18$~\cite{Sidorov}
imply that B-doped diamond is close to or in the crossover regime. 
 In general, the localization of single-particle wave functions leads to 
a small coherence length and a low superfluid density. 
 Therefore, a high upper critical field $H_{\rm c2}$  
and a large GL parameter $\kappa$ are expected in the crossover regime. 
 The localization effect revealed by electric resistivity 
measurements~\cite{Takano,Ishizaka}  
also indicates that B-doped diamond is close to the crossover regime.

 These observations should be contrasted to those of B-doped SiC. 
 B-doped SiC is a type I superconductor whose critical field $H_{\rm c}$ 
is 10$^{-3}$ times less than the upper critical field of 
B-doped diamond.~\cite{Ren} 
 Its GL parameter $\kappa \sim 0.35$~\cite{Ren,Kriener} is much less than that 
 of B-doped diamond. 
 A small residual resistivity implies that the effect of Anderson localization 
is negligible for superconductivity in B-doped SiC.  
 These marked differences between diamond and SiC indicate that 
B-doped SiC is in the host band regime far from the crossover regime. 
 This is consistent with the acceptor level of boron atoms 
in the SiC $\sim 0.3$ eV,~\cite{Muranaka,acceptorlevelSiC} 
which is smaller than that in diamond. 
 It is expected that the B-doped Si is far from the crossover regime 
because its acceptor level of $\sim 0.045$ eV~\cite{Muranaka,acceptorlevelSi} 
is the smallest among those of diamond, SiC, and Si. 
 We have illustrated these interpretations in Fig.~8.

 Before closing the section, we will comment on the difference 
between B-doped SiC and Al-doped SiC. 
 Both compounds have similar $T_{\rm c}$ values of $\sim 1.5$ K. 
 However, Al-doped SiC seems to be a type II superconductor 
with $\kappa \sim 1.8$ in contrast to B-doped SiC being a type I 
superconductor. 
 Since the acceptor level of Al-doped SiC of $\sim 0.2$ eV is smaller than 
that of B-doped SiC, the difference in the GL parameter $\kappa$ 
seems to be incompatible with the interpretation based on Fig.~8. 
 However, the difference between B-doped SiC and Al-doped SiC 
may be understood in the following way. 
 We expect that silicon sites play more important roles 
in the superconductivity than carbon sites,  
because the DOS at the Fermi level mainly arises from 
silicon sites.~\cite{Muranaka} 
 It has been pointed out that boron atoms mainly substitute for 
carbon atoms.~\cite{Muranaka} 
 If aluminium atoms substitute for silicon atoms rather than for carbon atoms, 
the random potential due to substitutional aluminum acceptors will 
affect superconductivity more significantly than that due to boron acceptors. 
 Then, type II superconductivity can occur in Al-doped SiC.  
 Further studies are desired to examine this possibility.

\section{Similarity to High-\Tc Cuprates}

 We point out the close relationship between doped semiconductors 
and high-\Tc cuprates. 
 We see the similarity between Fig.~8 and the typical phase diagram of 
high-\Tc cuprates. 
 $s$-Wave superconductivity and 
Anderson localization in doped semiconductors correspond to  
$d$-wave superconductivity and the 
antiferromagnetic Mott insulator in high-\Tc cuprates, respectively.  
 The host band and impurity band regimes in the former correspond to the 
overdoped and underdoped regimes in the latter, respectively. 

 We have claimed that the pseudogap in the under-doped high-\Tc cuprates 
is a precursor of superconductivity.~\cite{Janko,YanaseReview,Yanase1999} 
 The pseudogap due to the same mechanism has been discussed 
for the doped semiconductors in this paper. 
 The same mechanism of the pseudogap has also been proposed 
to high-density quark matters.~\cite{Kitazawa} 
 Note that the pseudogap in these systems does not arise from  
the shift in chemical potential, which describes the BCS-BEC crossover 
in zeroth-order approximation.~\cite{Eagles,Leggett,Nozieres} 
 In general, the shift in chemical potential is quantitatively important 
in low-density systems, such as trapped cold fermion gases,~\cite{Stringari}  
while higher-order corrections, namely, resonance scattering 
between incoherent Cooper pairs and quasiparticles, 
play an essential role in high-density systems.~\cite{YanaseReview} 
 Although a doped semiconductor seems to be a low-density system, 
its density is effectively high in the impurity band regime 
because the impurity band is nearly half-filled. 
 Actually, the shift in chemical potential is negligible in our calculation.  
 The BCS-BEC crossover described by the shift in chemical potential was first 
proposed for superconductivity in semiconductors.~\cite{Eagles} 
 However, this is not the case in our calculation.

 Many lines of experimental evidence have been obtained for  
microscopic inhomogeneity in high-\Tc cuprates.~\cite{Pan} 
 According to the results of RSTA, 
the microscopic inhomogeneity generally appears in small-coherence-length 
non-$s$-wave superconductors.~\cite{Yanase2006} 
 In contrast to that of the $s$-wave superconductor discussed in this paper, 
the microscopic inhomogeneity of non-$s$-wave superconductors 
is not necessarily accompanied by the localization of quasiparticles. 
 A weak disorder leads to microscopic inhomogeneity in the non-$s$-wave 
superconductor with a small coherence length. 
 On the other hand, the $s$-wave superconductivity is robust 
against nonmagnetic disorders in accordance with the Anderson 
theorem~\cite{Anderson-imp} until Anderson localization occurs.

 We note that the microscopic inhomogeneity of high-\Tc cuprates 
is enhanced by the multi-criticality of antiferromagnetism and 
$d$-wave superconductivity.~\cite{Yanase2007} 
 An antiferromagnetic correlation develops in the ``dirty region'', 
while a local superconductivity occurs in the ``clean region''. 
 It is expected that the insulating state near the SIT in high-\Tc cuprates 
is neither the simple Mott insulator nor the Anderson insulator in analogy 
with the doped semiconductors discussed in \S5.5. 
 It is expected that the localized Cooper pairing state is realized  
in a variety of systems near the SIT.

\section{Summary and Discussion}

 We have investigated localization and superconductivity 
in doped semiconductors on the basis of the disordered attractive 
Hubbard model in three dimensions. 
 We focused on the crossover from the superconductivity in 
the host band to that in the impurity band. 
 Anderson localization is accompanied by the crossover. 
 We found that the effective interaction for the paring is enhanced 
in the crossover regime. 
 The superconducting correlation is enhanced in the crossover regime, 
but suppressed by the mesoscopic and thermal fluctuations in superconductivity 
with approach to the impurity band regime. 
 The dome shape of the superconducting state is expected in the phase diagram 
against the impurity potential \Ui and the temperature $T$. 
 The long-range coherence is destroyed in the impurity band regime, 
and then the short-range correlation leads to a pseudogap. 
 The insulating state due to the localization of Cooper pairs 
is expected to be realized in the highly disordered regime.

 We proposed that the marked differences between B-doped diamond, 
SiC, and Si are understood by taking into account the impurity potential 
of these compounds, 
which are determined by the acceptor level in the lightly doped region. 
 B-doped diamond is close to or in the crossover regime, 
while the others are in the host band regime far away from the crossover regime. 
 The experimental results on transition temperature, 
critical magnetic field, residual resistivity, and 
the feature of the SIT are consistent with our interpretation.

 Two mechanisms have been proposed for the SIT in $s$-wave superconductors. 
 One is the so-called fermionic effect, which arises from the long-range 
Coulomb interaction~\cite{Maekawa,Kuroda,Finkelstein,Ikeda}. 
 The other is the bosonic effect, namely, the fluctuations 
in superconductivity~\cite{Fisher,Ghosal,Feigelman,Dubi,
Vinokur,Shklovskii2007,Beloborodov}. 
 Although the SIT in some bulk materials can be understood by taking the 
fermionic effect into account,~\cite{Belitz} 
the bosonic effect has been indicated by the recent experiments 
on disordered thin films,  such as Bi, InO$_x$, and TiN films.~\cite{Sambandamurthy,Gantmakher,Crane,Baturina,SacepeTiN,SteinerInO,Sarwa,
Vinokur,Spathis,Parendo,Stewart} 
 We have pointed out that a significant bosonic effect appears 
in a doped wide-gap semiconductor,  which may be realized 
in B-doped diamond.

 In summary, we have investigated superconductivity and localization 
with focus on the crossover from the host band regime 
to the impurity band regime in doped semiconductors. 
 The superconducting correlation is enhanced in the crossover regime 
where the electrons have a dual nature, namely, the itinerant and 
localized characters. 
 Anderson localization in the crossover regime is accompanied by  
the mesoscopic and thermal superconducting fluctuations in  
superconductivity. 
 The similarity of doped semiconductors to high-\Tc cuprates and 
disordered thin films has been discussed.

\section*{Acknowledgments}

 The authors are grateful to 
J. Akimitsu, N. Hayashi, K. Ishizaka, M. Kriener, Y. Maeno, H. Mukuda, 
T. Muranaka, J. Nakamura, N. Nishida, T. Nishizaki, Y. Ohta, T. Shirakawa, 
Y. Takano, T. Wakita, and T. Yokoya for valuable discussions.  
 This study was financially supported by the Nishina Memorial 
Foundation and Grants-in-Aid for Young Scientists (B) and 
for Scientific Research on Priority Areas (No. 17071002) 
from MEXT, Japan. 
 Numerical computation in this work was partly carried out 
at the Yukawa Institute Computer Facility.

\end{document}